\documentclass{PoS}

\newcommand{\der}{\mathrm{d}}

\newcommand{\mev}{\ensuremath{\mathrm{MeV}}}
\newcommand{\gev}{\ensuremath{\mathrm{GeV}}}

\newcommand{\tev}{\ensuremath{\mathrm{TeV}}}

\newcommand{\MET}{\mbox{\ensuremath{E \kern-0.6em\slash_{\rm T}}}}
\newcommand{\MHT}{\mbox{\ensuremath{H \kern-0.75em\slash_{\rm T}}}}
\newcommand{\ellell}{{\ensuremath{\ell^+\ell^-}}}
\newcommand{\ee}{{\ensuremath{e^+e^-}}}
\newcommand{\mumu}{{\ensuremath{\mu^+\mu^-}}}
\newcommand{\tautau}{{\ensuremath{\tau^+\tau^-}}}
\newcommand{\ra}{\ensuremath{\rightarrow}}
\newcommand{\pb}{\ensuremath{\mathrm{pb}}}

\newcommand{\pbi}{\ensuremath{\mathrm{pb}^{-1}}}
\newcommand{\fbi}{\ensuremath{\mathrm{fb}^{-1}}}

\title{Electroweak physics at the Tevatron}

\ShortTitle{Electroweak physics at the Tevatron}

\author{\speaker{Thomas NUNNEMANN}%
  \thanks{for the D\O{} and CDF Collaborations}\\
        Ludwig-Maximillians Universit\"{a}t, Munich, Germany\\
        E-mail: \email{Thomas.Nunnemann@lmu.de}}

\abstract{
Recent measurements of processes involving the electroweak bosons, $Z$ and $W$,
performed at the Fermilab Tevatron Collider are summarized. The large
integrated luminosities collected by both the D0 and CDF Collaborations enable
precise measurements of differential single boson production cross sections,
the determination of the $W$ mass with unprecedented precision, and the
observation of diboson production including $ZZ$. 
}

\FullConference{Physics at LHC 2008\\
		 29 September - October 4, 2008\\
		 Split, Croatia}

\begin{document}

\section{Introduction}

The analysis of single and pair production of the weak vector bosons 
$Z$ and $W$ in hadron-hadron collisions provides not only insight into the 
electroweak interaction (e.g. through the measurements of the couplings of
quarks and leptons to the $Z$ boson) but also allows to study the strong 
interaction (e.g. with measurements of differential boson production cross 
sections) or even to possibly detect new phenomena beyond the Standard
Model (SM, e.g. via an observation of anomalous triple-gauge couplings).
Furthermore, a precise measurement of the $W$ boson mass $M_W$ provides an 
indirect constraint on the mass of the long-sought Higgs boson given the
radiative corrections to $M_W$. 

In $p\bar{p}$ collisions at the Fermilab
Tevatron collider, the weak boson decays in leptonic final states exhibit 
very clear experimental signatures. In fall 2008, both the CDF and D0
experiments have collected data sets with integrated luminosities beyond
4\,\fbi{} at the collision energy of $\sqrt{s} = 1.96\,\tev$.
In the following, we summarize recent results of both collaborations on 
differential single boson production cross sections, on the mass of the $W$
boson and on diboson production.

\section{Single boson production}
Inclusive cross sections for the production of $Z$ and $W$ bosons decaying
into final states with $e$, $\mu$ or $\tau$ were previously published or 
presented as preliminary results~\cite{CDFweb,D0web}.
The D0 collaboration has recently updated their measurement of the inclusive
$Z$ boson production cross section in the 
$Z\rightarrow \tau \tau$ decay channel using a $1\,\fbi$ data 
set collected with an inclusive muon trigger~\cite{Abazov:2008ff}. 
In this analysis, one of the two $\tau$ candidates is reconstructed as a 
decay muon whereas the second is identified as a hadronically or 
electronically decaying $\tau$ using neural networks. 
The measurement yields
$\sigma_{Z} \cdot Br(Z\ra \tautau) =
237 \pm 15 \mathrm{(stat)}
\pm 18 \mathrm{(syst)} \pm 15 \mathrm{(lum)}\,\pb$,
which is in good agreement with the SM prediction. 
Furthermore, it verifies the
capability to identify isolated $\tau$ leptons, which is critical
in the search for the Higgs boson in supersymmetric extensions of the SM.

The large data sets being accumulated by both experiments enable precise
measurements of differential weak boson production cross sections, which allow
to set additional constraints on parton distribution functions (PDFs) and to
test corrections to the lowest-order predictions of 
Quantum Chromodynamics (QCD).

\subsection{$Z$ boson transverse momentum distribution}
QCD corrections to vector
boson production manifest themselves in the radiation of additional
quarks or gluons in the final state which gives rise 
not only to the production of associated jets, but also to a
substantial transverse momentum $p_T$ of the produced boson.
At large $p_T$, fixed-order calculations in perturbative QCD (pQCD)
are applicable and
have been derived up to next-to-next-to-leading order 
(NNLO)~\cite{Melnikov:2006kv}. For low $p_T$, where the emission of multiple
soft gluons becomes important, the leading logarithms in the perturbative
expansion can be resummed. The Monte Carlo generator {\sc resbos} accounts 
for additional non-perturbative corrections using a 
form factor~\cite{Landry:2002ix}.

The D0 collaboration has recently published a measurement of the $p_T$
distribution in $Z/\gamma^*\ra \ee$ events based on a data set with 
1\,\fbi~\cite{Abazov:2007nt}. 
The NNLO pQCD calculation is found to describe the shape of the distribution 
at $p_T>30\,\gev$, but underestimates the measured rate by $\sim 25\%$.
The data in the low $p_T$ region is well modeled by {\sc resbos} and 
disfavors a recently suggested modification of the form factor (small-$x$
broadening)~\cite{Berge:2004nt}.

The D0 collaboration has also presented a preliminary measurement of $g_2$,
a parameter in the non-perturbative form factor, using a data set of 2\,\fbi{}
and both $\ee$ and $\mumu$ decay signatures~\cite{D05755}. A new method
with a reduced sensitivity to the lepton $p_T$ resolution is applied, which
yields a measurement of $g_2$ as precise as the previous world average.
Future measurements of $M_W$ would benefit from an increased precision on the
form factor as the systematic uncertainty due to the modeling of weak boson
production would decrease.

\subsection{$Z$ boson rapidity distribution}
At leading order the boson rapidity $y$ is directly related to
the momentum fraction of the scattering partons 
$x_{1,2} = M_Z/\sqrt{s} \cdot e^{\pm y}$. Thus, its distribution at
large $|y|$ probes the PDFs at high momentum
transfer $Q^2 \approx M_Z^2$ and at both very large and low $x$.  
Similar to a previous D0 
publication~\cite{Abazov:2007jy} a preliminary measurement by the CDF
collaboration~\cite{CDFweb} finds a good agreement with higher-order pQCD.
The data is compared to various predictions using different PDF sets, but its
current precision does not provide significant additional constraints on PDFs.

\subsection{$W$ boson charge asymmetry}
As $u$ quarks carry on average a higher momentum fraction $x$ than $d$ quarks,
$W^+$ ($W^-$) bosons are preferentially boosted along the $p$ ($\bar{p}$)
direction, thus resulting into a charge asymmetry 
\[
A(y_W) =\frac{\der\sigma^+/\der y_W - \der\sigma^-/\der y_W}{\der\sigma^+/\der y_W + \der\sigma^-\der y_W}
\]
in the $W$ boson rapidity 
distribution. In leptonic $W$ decays $y_W$ cannot be directly determined since
the longitudinal momentum of the decay neutrino is unmeasured. However,
the pseudorapidity $\eta_\ell$ of the charged lepton from the $W$ decay is  
correlated with $y_W$ given the $V-A$ coupling of the weak interaction.
In addition, for the lepton asymmetry
\[
A(\eta_\ell) = \frac{\der\sigma(\ell^+)/\der\eta_\ell - \der\sigma(\ell^-)/\der\eta_\ell}{\der\sigma(\ell^+)/\der\eta_\ell + \der\sigma(\ell^-)/\der\eta_\ell}
\]
the systematic uncertainty related to lepton reconstruction largely cancels.

\begin{figure}
\includegraphics[width=.46\textwidth]{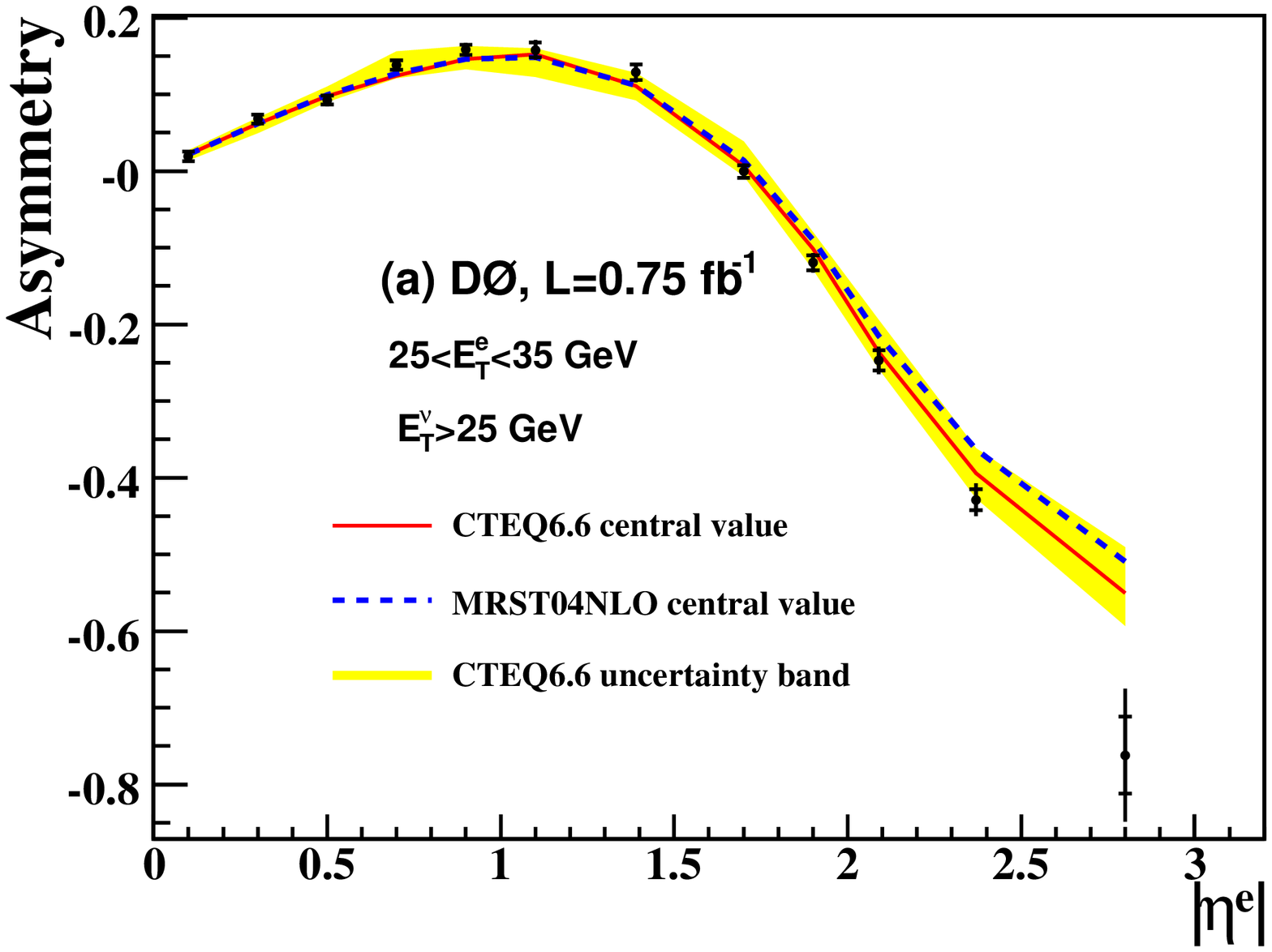}
\hspace{0.08\textwidth}\includegraphics[width=.46\textwidth]{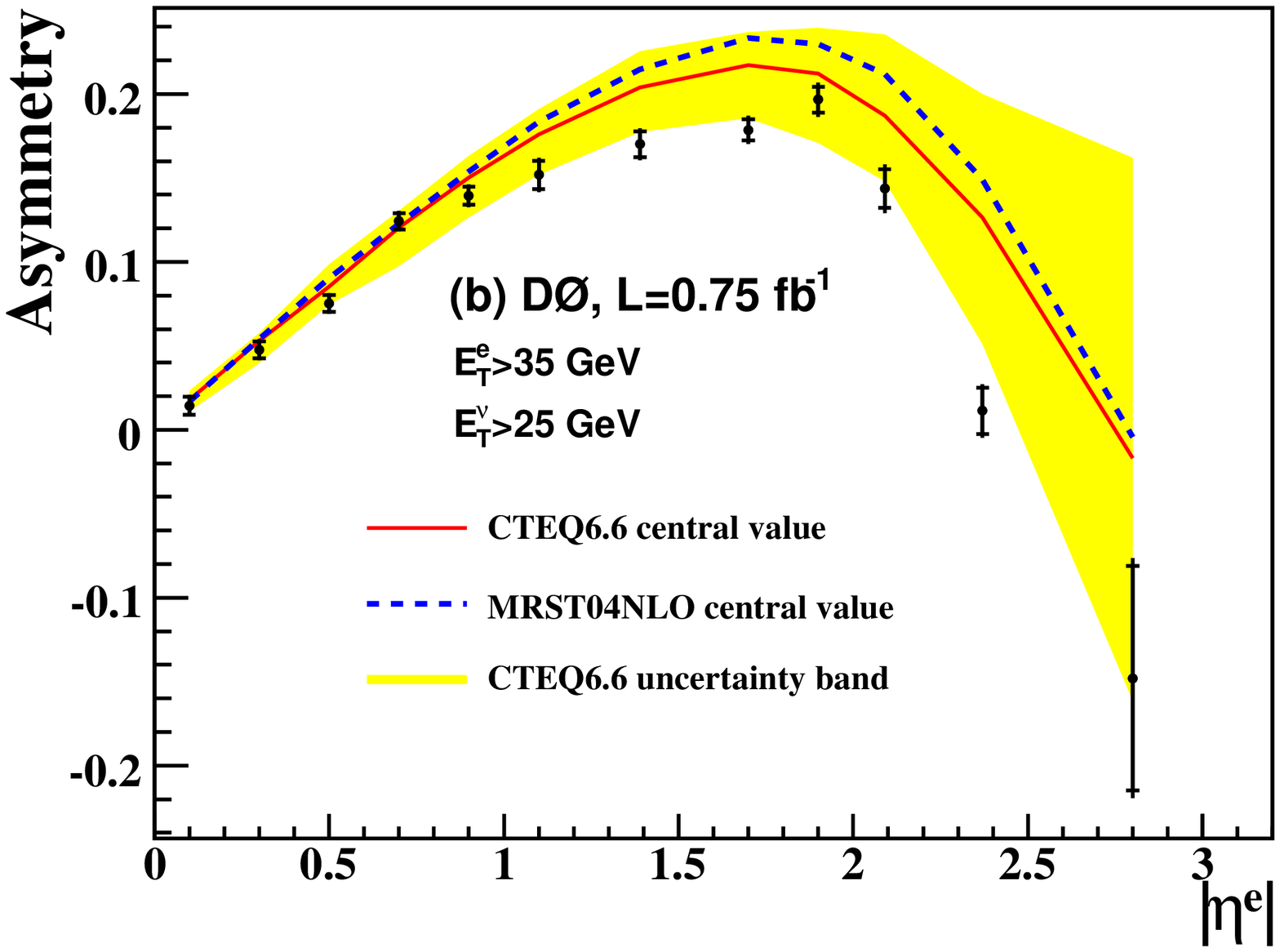}
\caption{The electron charge asymmetry distribution $A(\eta_e)$ in two 
$E_T^e$ bins: $25\,\gev < E_T^e < 35\,\gev$ for (a) and  $E_T^e > 35\,\gev$
for (b) compared to the predictions based on the CTEQ6.6 
(including uncertainty band) and MRST04NLO PDF sets~\cite{Abazov:2008qv}.}
\label{fig1}
\end{figure}

Fig.~\ref{fig1} shows $A(\eta_e)$ as measured by the D0 collaboration in two
bins of electron transverse energy $E_T^e$, which probe different regions of
$y_W$ (and thus $x$) for fixed $\eta_e$~\cite{Abazov:2008qv}.
For nearly all data points and in
particular for $E_T^e>35\,\gev$ the experimental uncertainties are 
significantly smaller that the PDF uncertainty band on the theoretical
prediction, which is calculated using the CTEQ6.6 error PDF 
sets~\cite{Nadolsky:2008zw}.

The CDF collaboration has developed a method to directly measure the $W$ boson
asymmetry $A(y_W)$ by reconstructing the $y_W$ distribution using a constraint
on $M_W$~\cite{CDF8942}. The two possible solutions for the neutrino momentum
are weighted with a probability which is iteratively 
determined from simulation.
This new method has an improved statistical sensitivity compared to 
the measurement of $A(\eta_\ell)$, but since for fixed values of $y_{W^\pm}$
the corresponding lepton pseudorapidities $\eta_\ell^+$ and $\eta_\ell^-$ 
are distinct, the geometric acceptances to reconstruct $W^+$ or 
$W^-$ bosons, respectively, can differ by large amounts and the systematic
uncertainty due to the lepton reconstruction is increased.
Similar to the D0 result on $A(\eta_e)$, CDF's preliminary measurement of 
$A(y_W)$ based on $W\ra e\nu$ decays and an integrated luminosity of $1\,\fbi$
has a higher precision than the PDF uncertainty 
on the prediction~\cite{CDF8942}.

\subsection{Forward-backward charge asymmetry in 
$Z/\gamma^*\rightarrow e^+ e^-$ events}

A measurement of the forward-backward charge asymmetry $A_{FB}$ as function of
the dilepton mass $M_\ellell$ in 
$Z/\gamma^*\rightarrow \ell^+ \ell^-$ events probes the interference between 
the
$\gamma^*$ and $Z$ propagators and is sensitive to the vector and 
axial-vector couplings $g_V^{u,d,\ell}$ and $g_A^{u,d,\ell}$ of the 
$u$ and $d$ quarks and the final-state lepton to the $Z$ boson.
Therefore, $A_{FB}$ also constrains the effective weak mixing angle
$\sin^2\theta_W^{\mathrm{eff}}$ and
a hypothetical heavy $Z^\prime$ boson would alter $A_{FB}$ for 
$M_\ellell \approx M_{Z^\prime}$.

A previous measurement of $A_{FB}$ based on
only $72\,\pbi$ of integrated luminosity performed by the CDF collaboration
put constraints on $g_V^{u,d}$
and  $g_A^{u,d}$ complementary to the results from LEP and 
HERA~\cite{Acosta:2004wq}. The D0 collaboration recently published a 
measurement of $A_{FB}$ up to dielectron masses $M_{ee}>300\,\gev$ using their
$1\,\fbi$ data set and extracted $\sin^2\theta_W^{\mathrm{eff}}$ with
a precision comparable to that of the LEP measurements of the inclusive
hadronic charge asymmetry~\cite{Abazov:2008xq}.

\section{Measurement of the $W$ boson mass and width}

\begin{figure}
\includegraphics[width=.46\textwidth]{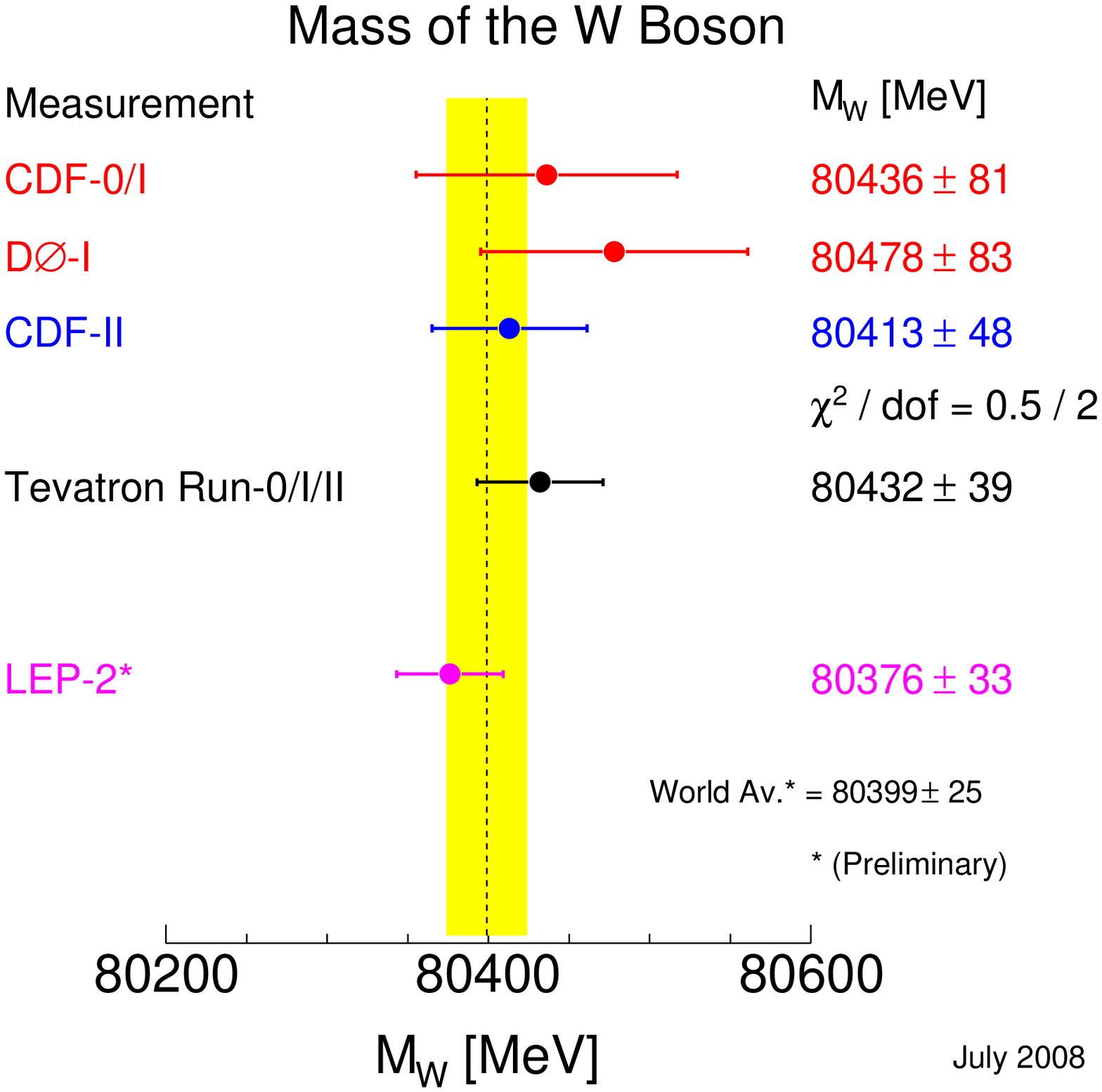}
\hspace{0.08\textwidth}\includegraphics[width=.46\textwidth]{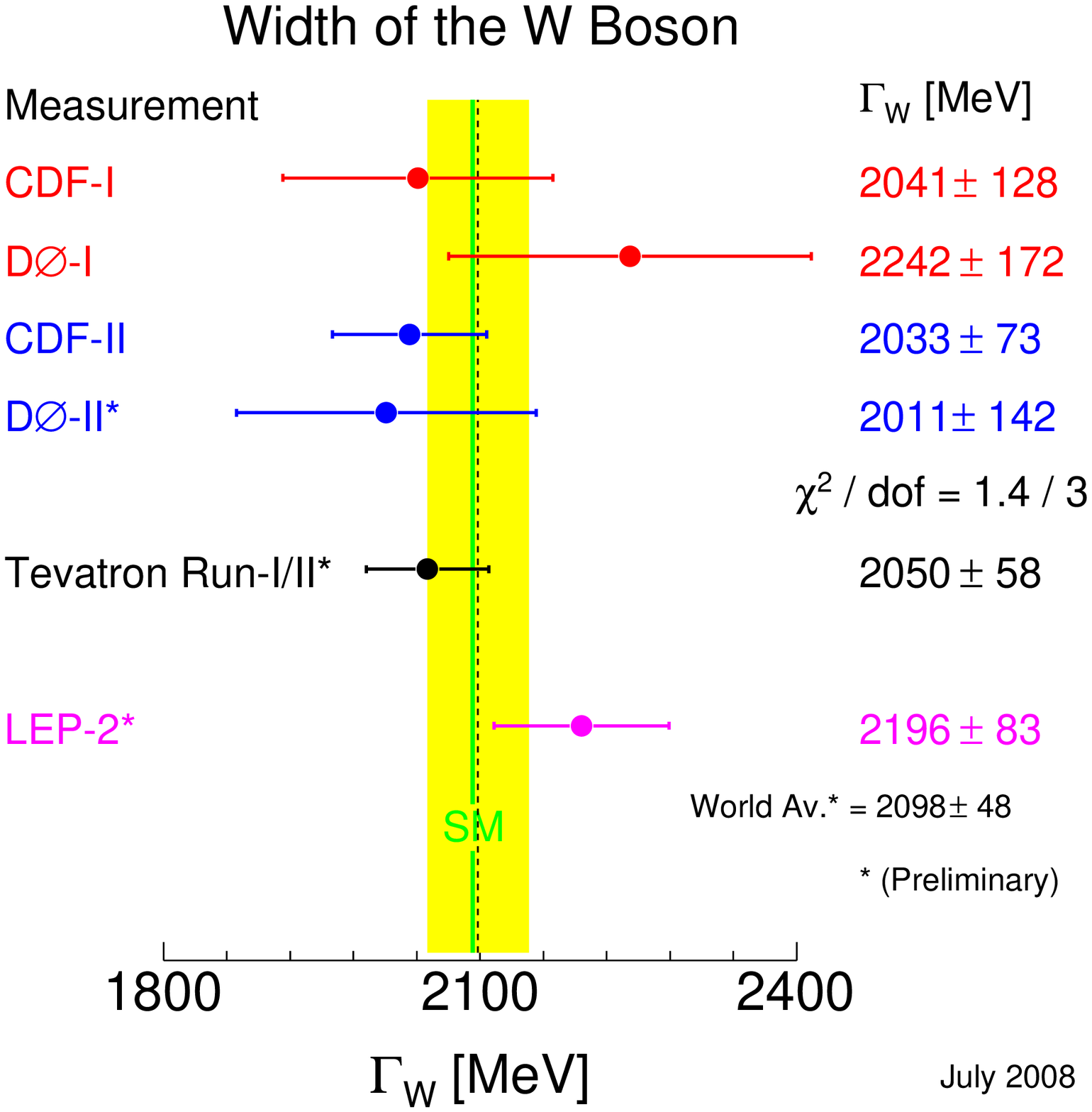}
\caption{Comparison of the measurements of the $W$ boson mass (left) and 
width (right) together with their averages~\cite{CDFD0:2008ut}.}
\label{fig2}
\end{figure}

The new combination of the measurements of the $W$ mass $M_W$ and width 
$\Gamma_W$ from the
Tevatron Runs and LEP~\cite{CDFD0:2008ut} is shown in Fig.~\ref{fig2}. The 
Run\,I results have been corrected to account for their outdated assumptions
on PDFs and $\Gamma_W$ or $M_W$, respectively.

The most precise single measurement of $M_W$ has been achieved by the CDF
collaboration using $W\ra e\nu$ and $W\ra \mu\nu$ events collected in a data
set with 200\,\pbi{} integrated luminosity~\cite{Aaltonen:2007ypa}.
For this result, the $W$ mass is derived from distributions of the 
transverse mass $m_T$, the
lepton $p_T$ and the missing transverse energy $\MET$. 
The statistical and systematic uncertainty contribute equally to
the precision on the measured $W$ boson mass $\Delta M_W = 48\,\mev$.

A preliminary CDF analysis based on a more than tenfold integrated
luminosity $L$ shows that the statistical uncertainty on $M_W$ scales
with approximately $\sqrt{L}$, which demonstrates that the measurement 
of $M_W$ does not
degrade with the larger energy pile-up 
in the calorimeter due to the increasing 
instantaneous luminosity~\cite{CDFweb}. Whereas the lepton energy scale of the
published measurement is mostly constrained by $J/\psi$ and $\Upsilon$
decays due to the limited number of $Z$ boson decays, the improved statistical
precision on $M_Z$ with increasing data sets is expected to significantly
reduce the scale uncertainty, which is one of the
dominating systematic errors in the measurement of $M_W$.

\section{Diboson production}

The pair production of the electroweak bosons, $W$, $Z$, and $\gamma$, probes 
the trilinear gauge boson
couplings predicted by the non-Abelian structure of the SM. Any deviation
from the expected couplings, commonly referred to as anomalous couplings,
would indicate the presence of new physics beyond the SM.
The Tevatron measurements are complementary to those performed at LEP, 
since the former probe higher energies and are sensitive to different
combinations of couplings. As a compilation of all diboson measurements of
both CDF and D0 collaborations can be found elsewhere~\cite{CDFweb,D0web},
only a few recent results will be presented below.

\subsection{Observation of $ZZ$ production}

\begin{figure}
\includegraphics[width=.4775\textwidth]{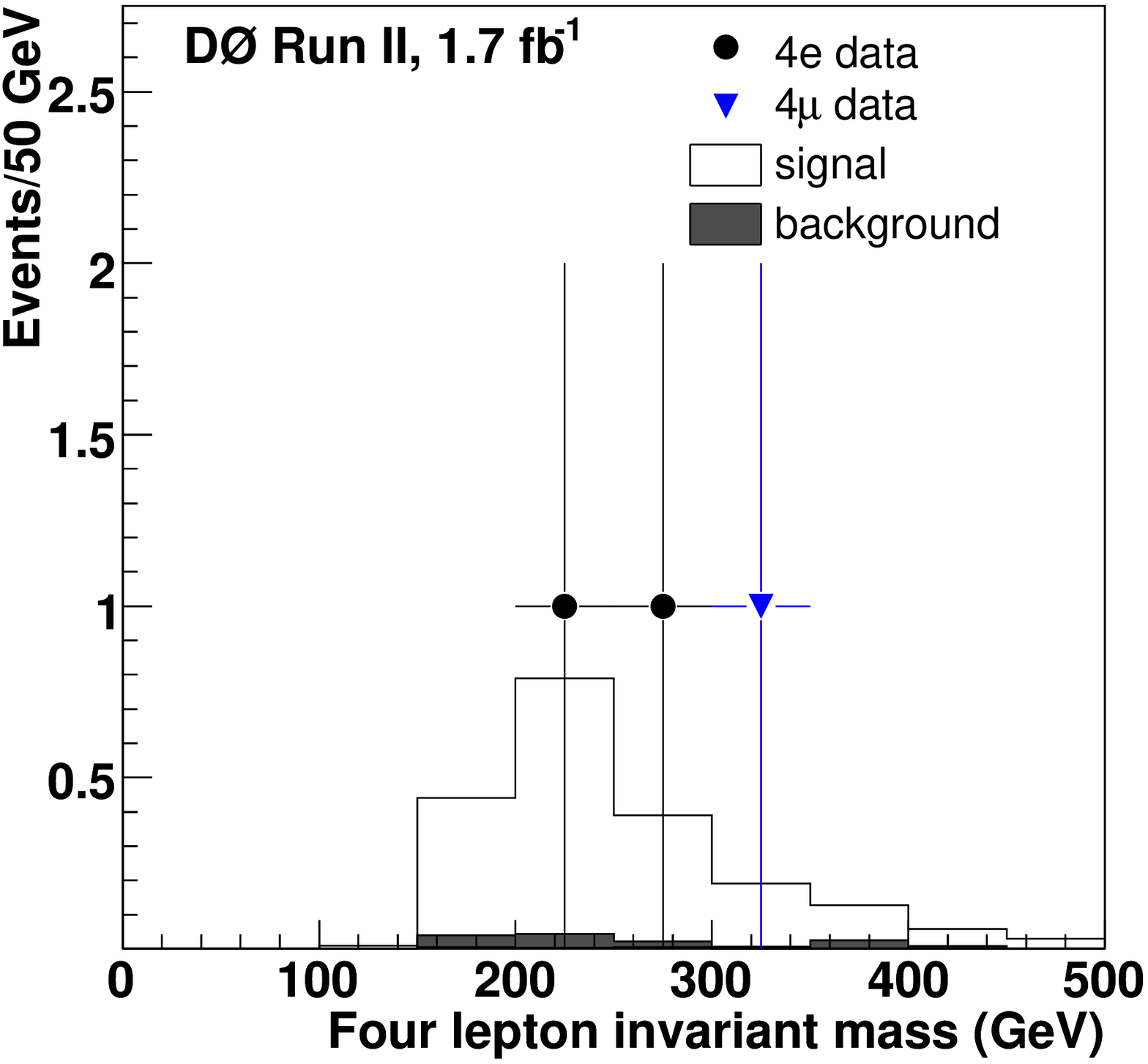}
\hspace{0.08\textwidth}\includegraphics[width=.4425\textwidth]{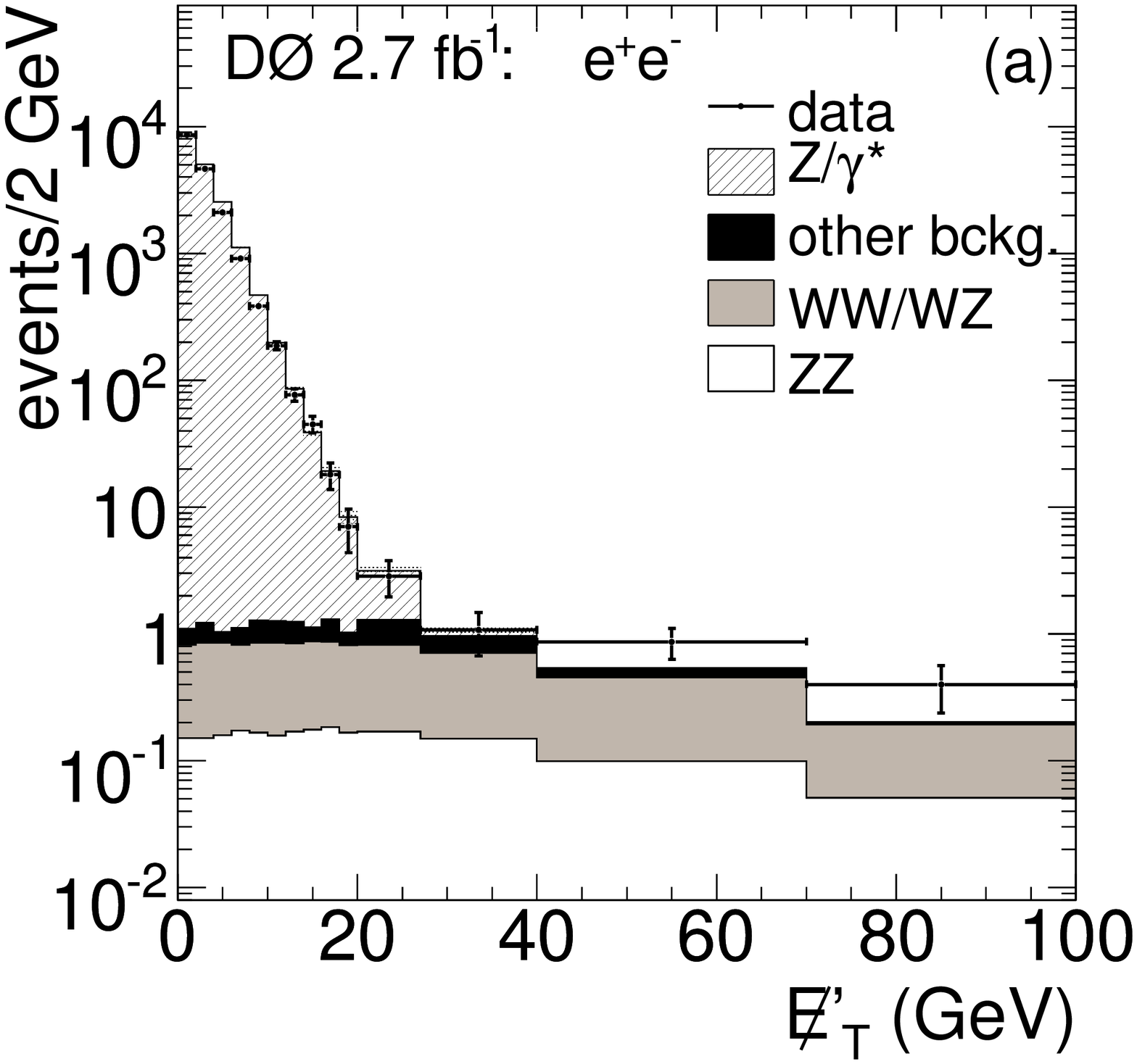}
\caption{Left: Distribution of four lepton invariant mass in the 
$ZZ\rightarrow \ell\ell\ell^\prime\ell^\prime$ selection~\cite{Abazov:2008gya}.
Right: Distribution of the missing transverse energy estimator $\MET^\prime$
in the $\rightarrow \ell\ell\nu\nu$ selection before applying the $\MET^\prime$
requirement~\cite{Abazov:2008yf}.}
\label{fig3}
\end{figure}

Following CDF's measurement of $ZZ$ production with a significance of 
$4.4\sigma$~\cite{Aaltonen:2008mv}, the D0 collaboration has reported
the first observation of this process with a significance of 
$5.7\sigma$~\cite{Abazov:2008gya}.
Two selections are applied. For $ZZ\rightarrow \ell\ell\ell^\prime\ell^\prime$
($\ell,\ell^\prime = e$ or $\mu$) production,
three candidate events, with an expected 
background of 0.14 events are found in 1.7\,\fbi{} of data
corresponding to a significance of $5.3\sigma$.
Fig.~\ref{fig3} shows the four lepton invariant mass for these events compared
to the expected signal and background distributions.
For the other channel, $ZZ\rightarrow \ell\ell\ell\nu\nu$, a new estimator
for the missing transverse energy with a reduced sensitivity to
instrumental mismeasurements resulting in an improved background rejection
has been developed (Fig.~\ref{fig3})~\cite{Abazov:2008yf}. After the 
signal is discriminated from the dominating $WW$ background using a likelihood
a significance of $2.7\sigma$ is observed.
The combined $ZZ$ production cross section is
$\sigma = 1.60\pm 0.63\mathrm{(stat)}^{+0.16}_{-0.17}\mathrm{(syst)}\,\pb$,
consistent with the SM prediction of $1.4\pm 0.1$\,pb derived at NLO.

\subsection{Limits on anomalous triple gauge couplings}

\begin{figure}
\includegraphics[width=.47\textwidth]{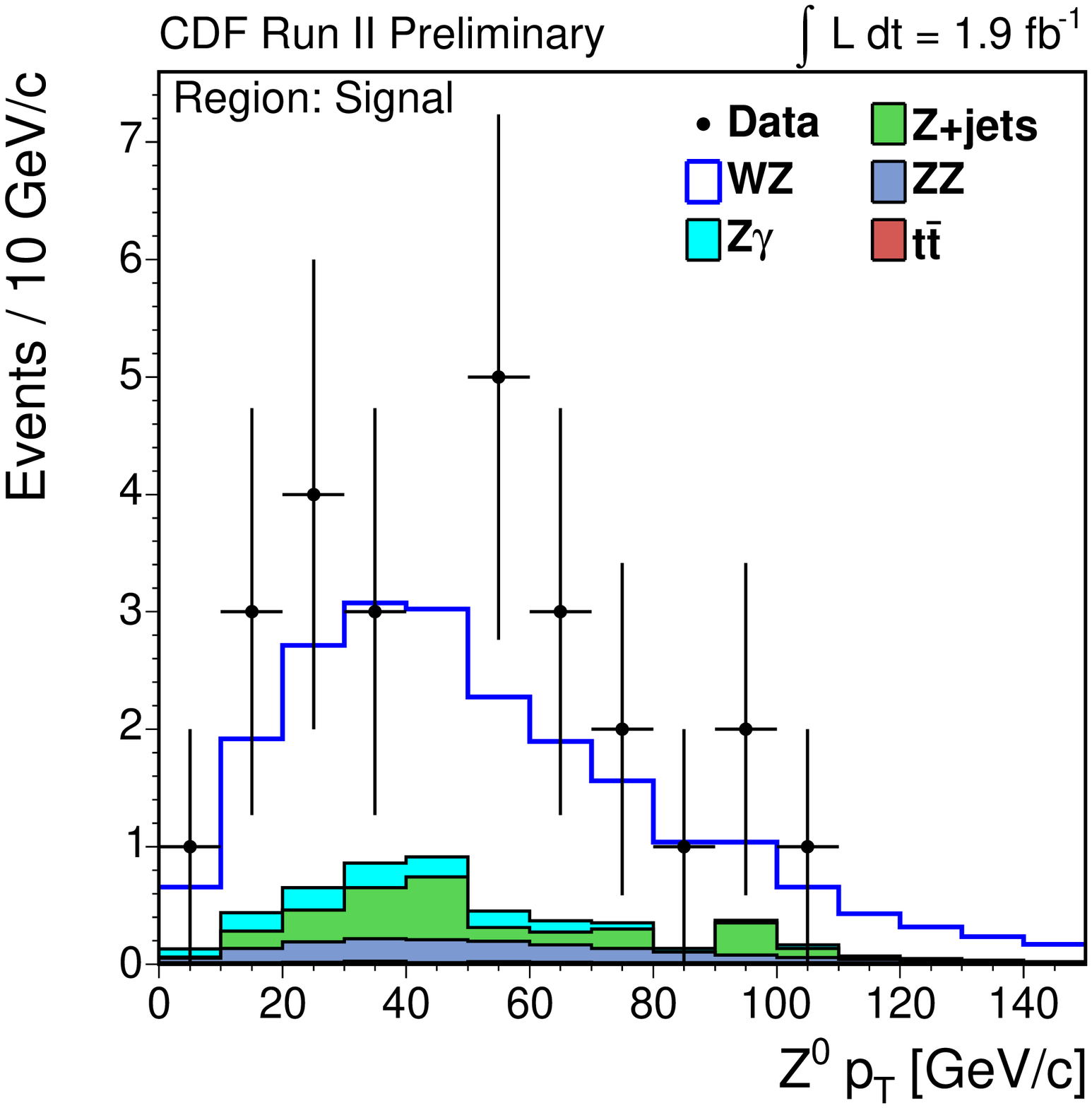}
\hspace{0.08\textwidth}\includegraphics[width=.45\textwidth]{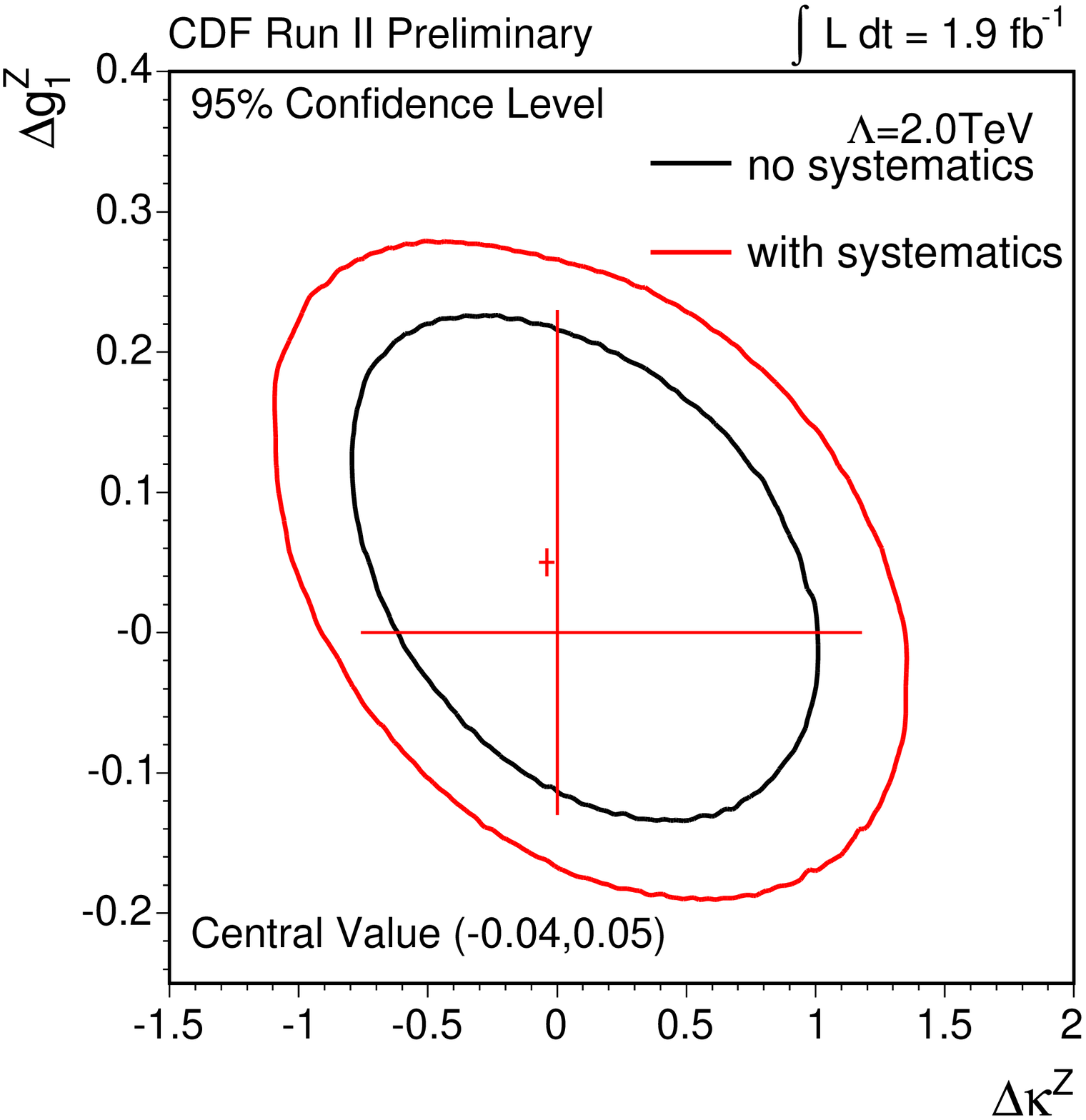}
\caption{Left: The $p_T$ distribution of the $Z$ boson in $WZ$ candidate 
events. Right: The one and two-dimensional limits on $\Delta g_1^Z$ and 
$\Delta\kappa_Z$ (deviations from SM values) derived from the distribution
shown on the left side~\cite{CDFweb}.}
\label{fig4}
\end{figure}

The analysis of $WZ$ production allows to study the $WWZ$ triple gauge coupling
independently of the $WW\gamma$ vertex contribution. For the $WWZ$ vertex
three CP conserving coupling parameters are defined, which take the following
values in the SM: $g_1^Z=1$, $\kappa_Z=1$, $\lambda_Z=0$.
In a recent preliminary measurement the CDF collaboration utilizes the $p_T$
distribution of $Z$ bosons in $WZ$ events (Fig.~\ref{fig4}) to constrain
anomalous contributions to the $WWZ$ triple gauge coupling~\cite{CDFweb}. 
As an example, the one and two-dimensional limits
on $g_1^Z$ and $\kappa_Z$ are shown in Fig.~\ref{fig4}.

\section{Conclusions}
The increasing luminosities at Tevatron enable the precise study of
single and diboson production as well as the measurement of fundamental
parameters like $M_W$. Furthermore precise measurements are expected from
the continuously increasing data sets which are collected by both the
CDF and D0 experiments.

\section*{Acknowledgments}
I would like to thank my colleagues from the CDF and D0 collaborations for
providing their excellent results and the organizers for the stimulating
conference.

\end{document}